\documentclass[%
 reprint,
 amsmath,amssymb,
 aps,
]{revtex4-2}

\usepackage[table,xcdraw]{xcolor}
\usepackage{graphicx}
\usepackage{dcolumn}
\usepackage{bm}

\usepackage{amsmath}
\usepackage{booktabs}
\usepackage{color}
\usepackage{hyperref}

\hypersetup{
  colorlinks=true,
  urlcolor=blue,
  linkcolor=red,
  citecolor=blue
}

\usepackage{multirow}
\usepackage{xfrac}
\usepackage{array, makecell}
\newcolumntype{C}[1]{>{\centering\arraybackslash}p{#1}}

\newcommand{\mnras}{Monthly Not. Roy. Astron. Soc.}

\newcommand {\ee} {\end {equation}}
\newcommand {\be} {\begin {equation}}

\newcommand{\cas}{\textstyle\frac}

\def\d{{\rm d}}
\def\e{{\rm e}}
\def\i{{\rm i}}

\newcommand{\bE}{{\mbox{\boldmath $E$}}}

\newcommand{\brh}{{\bar H}}

\newcommand{\bv}{{\mbox{\boldmath $v$}}}

\DeclareFontFamily{U}{mathb}{\hyphenchar\font45}
\DeclareFontShape{U}{mathb}{m}{n}{
      <5> <6> <7> <8> <9> <10> gen * mathb
      <10.95> mathb10 <12> <14.4> <17.28> <20.74> <24.88> mathb12
      }{}
\DeclareSymbolFont{mathb}{U}{mathb}{m}{n}
\DeclareFontSubstitution{U}{mathb}{m}{n}

\DeclareMathSymbol{\curvearrowtopleft} {\mathrel}{mathb}{"F0}
\DeclareMathSymbol{\curvearrowbotright}{\mathrel}{mathb}{"F4}
\DeclareMathSymbol{\curvearrowtopright}{\mathrel}{mathb}{"F1}

\usepackage{dcolumn}
\usepackage{bm}

\begin{document}

\preprint{APS/123-QED}

\title{Eigenvalues and Eigenfunctions of Landau Damping Oscillations \\in Very Weakly Collisional Plasma} 

\author{Evgeny V. Polyachenko}
\affiliation{Institute of Astronomy, Russian Academy of Sciences, 
48 Pyatnitskaya st, Moscow 119017, Russia}
\email{epolyach@inasan.ru}
\affiliation{SnT SEDAN, University of Luxembourg, 
29 boulevard JF Kennedy, L-1855 Luxembourg, Luxembourg}
\email{evgeny.polyachenko@uni.lu}

\author{Ilia G. Shukhman}
\affiliation{Institute of Solar-Terrestrial Physics, Russian Academy of Sciences, \\ Siberian Branch, P.O. Box 291, Irkutsk 664033, Russia}
\email{shukhman@iszf.irk.ru}

\date{\today}

\begin{abstract}
Landau-damped oscillations in collisionless plasmas, described by van Kampen and Case, are quasi-modes, representing a continuous superposition of singular eigenfunctions, not true eigenmodes. Recent work by Ng et al. shows that even rare collisions replace these singular modes with discrete regular modes having complex eigenvalues for the phase velocity (or frequency), approaching Landau eigenvalues in the collisionless limit. We analytically derive approximate expressions for the eigenvalue correction due to rare collisions and for the shape of the eigenfunction describing DF perturbations in velocity space, demonstrating its increasing oscillations in the resonance region as the collision frequency tends to zero. We also obtain approximate expressions for the resonance region's width and peak value, and the oscillation period within it. We validate these analytical results with high-precision numerical calculations using a standard linear matrix eigenvalue problem approach.
\end{abstract}

\pacs{}

\maketitle

\section{Introduction}

The initial-value problem of \citet{Lan_46} showed that small perturbations in a collisionless Maxwellian plasma exhibit exponential decay (Landau damping) of macroscopic quantities like density and electric field. However, as shown in the same work, the electron distribution function (DF) does not decay, but instead becomes increasingly fragmented in velocity space. This indicates that the exponential decay is not a true eigenoscillation with a complex frequency $\omega_{\rm L}$ with a negative imaginary part. In other words, it does not correspond to any eigenfunction (EF) for the DF that would evolve while preserving its shape in velocity space. Several years later, \citet{vK_55} and \citet{Case59} clarified that the spectrum of eigenmodes in a Maxwellian plasma is real and continuous, with frequencies $\omega$ (or phase velocities $\omega/k$) spanning an infinite range. The corresponding EFs of the DF, $f_c(v)$, are singular at $v=c$ and form a complete set. Landau damping is then understood as the evolution of a continuous superposition of these van Kampen modes. Thus, the perturbation associated with exponential Landau damping is a quasi-mode, not a true eigenmode.

Similar connections between damped oscillations and singular van Kampen modes exist in collisionless self-gravitating stellar media (see, e.g.,\cite{PSB,2021MNRAS.507.2241L})  and stable inviscid shear flows \cite{BalMor1995, PolShu2022, PS2024}. This unphysical behavior of the EFs arises from neglecting dissipative factors, such as collisions in the Boltzmann equation or viscosity in the Navier-Stokes equation. These factors become important as gradients of perturbed quantities increase, inevitably influencing the fine-scale structure of the DF or flow velocity profile, regardless of how small the collision frequency or viscosity coefficient is.

\citet{Escande_2018} used an original N-body approach in which collisions are intrinsic property of plasma. While their approach successfully reproduces the Landau damping rate of the electric field, it has a fundamental limitation for our purposes: their model's collisions can be chosen sufficiently rare to obtain the correct damping rate but remain too frequent for examining the detailed behavior of the DF in the collisionless limit. This makes their approach irrelevant for our goal of tracking DF
behavior as collision frequency approaches zero.

In contrast, \citet{Ng1999} showed that adding a collision term with a second derivative with respect to velocities to the collisionless Vlasov \cite{Vlasov45} kinetic equation provides a traditional approach that radically alters the eigenmode picture. The complex frequencies associated with Landau damping in the collisionless case become eigenfrequencies of true eigenmodes. These frequencies receive a small correction related to the collision frequency, transitioning smoothly to the Landau frequencies in the collisionless limit. The corresponding collisional EFs are regular. Furthermore, \citet{Ng2004} demonstrated that the spectrum of eigenmodes corresponding to the complex Landau frequencies is complete, replacing the real van Kampen mode spectrum, which vanishes with any non-zero collision frequency.

However, the limiting transition of regular collisional EFs as the collision frequency approaches zero remains unclear. While the transition of collisional eigenvalues to collisionless Landau values is smooth, tracing the EF transition requires extremely small dimensionless collision frequencies, hindering traditional numerical methods. 

In this work, using the most important least-damped mode as an example, we aim to trace the transformation of its characteristics as the collision frequency approaches zero. We derive approximate analytical expressions in Section II and validate them with high-precision numerical calculations in Section III. Section IV summarizes the results.

\section{Analytical Expressions}

Here, we derive approximate analytical expressions for eigenvalues and EFs with very rare collisions. Following \citet{Lan_46}, we consider potential plasma oscillations, $\bE=-\nabla \Phi$, assuming ions are immobile. For simplicity, we restrict ourselves to one-dimensional perturbations and start, as in \cite{Ng1999}, with the linearized Boltzmann equation (with a Lenard-Bernstein collision term \cite{LenBern}) and the Poisson equation:
\begin{align}
\frac{\partial\, \delta f}{\partial t}\!+\!v\,\frac{\partial\, \delta f}{\partial x}&\!+\!\frac{e}{m}\,\frac{\partial\,\delta\Phi}{\partial x}\, \frac{\partial f_0}{\partial v}=\nu\,\frac{\partial }{\partial v}\,\Bigl(v\,\delta f\!+\!\sigma^2\frac{\partial\,\delta f}{\partial v}\Bigr)\,,
\label{eq:Bol} \\[-1mm]
\frac{\partial E}{\partial x}&=-\frac{\partial^{\,2}\delta \Phi}{\partial x^2}=-4\pi e\! \int \d v\, \delta f \,,
\label{eq:Poisson}
\end{align}
where $\delta f(t,x,v)$ and $\delta\Phi(t,x)$ are the electron DF and electric potential perturbations, $-e$ and $m$ are the electron charge and mass, and $\nu$ is the collision frequency. The unperturbed DF is Maxwellian: $f_0(v)={n_0}\,(2\pi\sigma^2)^{-1/2}\,\exp\bigl[-{v^2}/({2\sigma^2})\bigr]$, where $n_0$ is the electron concentration and $\sigma$ is the thermal velocity. For perturbations $\delta f = f(v)\,\e^{\i (kx-\omega\,t)}$ and $\delta\Phi =\phi\,\e^{\i (kx-\omega\,t)}$, we switch to dimensionless variables, noting $-k^2\phi=4\pi e\int\d v\, f$. Dimensionless variables are $u={v}/({\sqrt{2}\,\sigma})$, $g(u)=\sqrt{2}\,\sigma\,{f}/{n_0}$, $g_0(u)=\sqrt{2}\,\sigma\,{f_0}/{n_0}=\e^{-u^2}/\sqrt{\pi}$, $\alpha=1/(kr_D)^2={4\pi n_0 e^2}/({m\,k^2\sigma^2})$, $c={\omega}/({\sqrt{2} k\,\sigma})$, and $\mu=\nu/(\sqrt{2} k\sigma)$, yielding
\begin{multline}
\hspace{-10pt}(u-c)\,g(u)-\eta(u) \hspace{-4pt} \int\limits_{-\infty}^\infty \hspace{-4pt}\d u'\, g(u') =
\frac\mu\i\frac{\d}{\d u}\,\Bigl(u\,g+\frac{1}{2}\,\frac{\d g}{\d u}\Bigr)\,,
\label{eq:Eq_for_g}
\end{multline}
where $\eta(u)\equiv ({\alpha}/{2})\,\d g_0/\d u=-(\alpha/\sqrt{\pi})\,u\,\exp(-u^2)$.

\subsection{Collisional Correction to Landau frequency}

For $\mu \ll 1$, a first-order correction in $\mu$ to the collisionless Landau eigenvalue can be found using perturbation theory. Starting with Eq. (\ref{eq:Eq_for_g}), we set
\begin{equation}
g(u)=g^{(0)}(u)+g^{(1)}(u)\,,\ \ c=c_{\rm L}+c_1\,,
\end{equation}
where $c_{\rm L}$ and $g^{(0)}(u)$ are the collisionless eigenfrequency and EF, satisfying
\begin{equation}
(u-c_{\rm L})\,g^{(0)}(u)-\eta(u)=0\,,\ \  g^{(0)}(u)=\frac{\eta(u)}{u-c_{\rm L}}\,,
\label{eq:g_0}
\end{equation}
with
\vspace{-6pt}
\begin{equation}
\int_{-\infty}^\infty \d u \, g^{(0)}(u)=1\,.
\label{eq:DEq}
\end{equation}

In a collisionless Maxwellian plasma, no EF exists on the real $u$-axis; the dispersion equation (\ref{eq:DEq}) has no solutions for any complex $c_{\rm L}$. However, an EF exists if $g^{(0)}(u)$ (\ref{eq:g_0}) is considered on a contour in the complex $u$-plane, passing below $c_{\rm L}$ \cite{PSB}. Therefore, the dispersion equation becomes:
\begin{equation}
-\frac{\alpha}{\sqrt{\pi}}\int\limits_\curvearrowbotright \d u\, \frac{u\,\e^{-u^2}}{u-c_{\rm L}}=1\,,
\label{eq:DEq1}
\end{equation}
where ``$\curvearrowbotright$'' indicates integration along a contour in the lower half-plane of the complex $u$-plane, passing below $u=c_{\rm L}$.
With this in mind, we obtain
\begin{multline}
    -c_1\,g^{(0)}(u)+ (u-c_{\rm L})\,g^{(1)}+\frac{\alpha}{\sqrt{\pi}}\,u\,\e^{-u^2}\int\limits_\curvearrowbotright \d u'\,g_1 (u')=\\
    -\i\mu\, \frac{\d}{\d u}\,\Bigl(u\,g^{(0)}+\frac{1}{2}\,\frac{\d g^{(0)}}{\d u}\Bigr)\,.
    \label{eq:for_c1}
\end{multline}
Dividing Eq.\,(\ref{eq:for_c1}) by $(u-c_{\rm L})$ and integrating over $u$ yields
\begin{multline}
    -c_1\int\limits_\curvearrowbotright \frac{g^{(0)}(u)\,du}{u-c_{\rm L}}+\int\limits_\curvearrowbotright g^{(1)}(u')\, du'\,\Bigl(1-
\int\limits_\curvearrowbotright g^{(0)}(u)\,du\Bigr)
=\\
-\i\mu \int\limits_\curvearrowbotright \frac{du}{u-c_{\rm L}}\frac{d}{du}\,\Bigl(u\,g^{(0)}+\frac{1}{2}\,\frac{d g^{(0)}}{du}\Bigr)
\end{multline}
The second term vanishes due to Eq.\,(\ref{eq:DEq1}). Thus, the solvability condition of  (\ref{eq:for_c1}) for $g^{(1)}$ requires
\begin{multline}
    \hspace{-8pt}c_1\int\limits_\curvearrowbotright\hspace{-2pt} \d u \frac {u\,\e^{-u^2}}
{(u-c_{\rm L})^2}=
\frac{c_{\rm L}\mu}{2\i}\int\limits_\curvearrowbotright
\hspace{-4pt}
\frac{\d u}{u-c_{\rm L}}\frac{\d}{\d u} \left[\frac{\e^{-u^2}}{(u-c_{\rm L})^2}\right],
\end{multline}
allowing us to find $c_1$. Integration by parts gives $c_1=-\frac{1}{2}\,\i\,\mu\,c_{\rm L}\,I_2/I_1$, where
\begin{equation}
I_1=\frac{\alpha}{\sqrt{\pi}}\int\limits_\curvearrowbotright \frac {u\,\e^{-u^2}\,\d u}
{(u-c_{\rm L})^2}, \ \ \ I_2= \frac{\alpha}{\sqrt{\pi}}\,\int\limits_\curvearrowbotright\
\frac{\e^{-u^2}\,\d u}{(u-c_{\rm L})^4}.
\end{equation}
For $I_1$, we have
\begin{multline}
I_1=\frac{\alpha}{\sqrt{\pi}}\int\limits_\curvearrowbotright \d u\, \frac{\e^{-u^2}}{u-c_{\rm L}}+c_{\rm L}\,\frac{\alpha}{\sqrt{\pi}}\int\limits_\curvearrowbotright \d u\, \frac{\e^{-u^2}}{(u-c_{\rm L})^2}\\
=J-2 c_{\rm L}\,\frac{\alpha}{\sqrt{\pi}}\int\limits_\curvearrowbotright \d u\, \frac{\e^{-u^2}\,u}{u-c_{\rm L}}=J+2c_{\rm L}\,,
\end{multline}
where we have used equality (\ref{eq:DEq1}), and defined
\begin{equation}
J=\frac{\alpha}{\sqrt{\pi}}\int\limits_\curvearrowbotright \d u\, \frac{\e^{-u^2}}{u-c_{\rm L}}\,.
\end{equation}
Integral $J$ can be found from the dispersion equation (\ref{eq:DEq1}),
\begin{equation}
-1=\frac{\alpha}{\sqrt{\pi}}\int\limits_\curvearrowbotright  \frac {\e^{-u^2}\,u\,\d u}
{u-c_{\rm L}}=\frac{\alpha}{\sqrt{\pi}}\int \e^{-u^2}\,\d u +c_{\rm L}\,J=\alpha +c_{\rm L}\,J,
\end{equation}
so $J=-(1+\alpha)/c_{\rm L}$. Thus,
\begin{equation}
I_1=\frac{2 c_{\rm L}^2-(1+\alpha)}{c_{\rm L}}\,.
\end{equation}

For $I_2$, three integrations by parts, followed by subsequent manipulations similar to those used for $I_1$, yield
\begin{equation}
I_2 =-\frac{4}{3}-\frac{2}{3}\,\bigl[(1+\alpha)-2 c_{\rm L}^2\bigr]=-\frac{4}{3}+\frac{2}{3}\,c_{\rm L}\,I_1\,. 
\end{equation}
Finally, we obtain
\begin{equation}
\Delta c\equiv  c_1=-\frac{\i\mu}{3}\,\left[1+\frac{2}{(1+\alpha)-2\,c_{\rm L}^2}\right]\,c_{\rm L}^2\,.
\label{eq:final_Omega_1}
\end{equation}
 Note that it can be verified that (\ref{eq:final_Omega_1}) is consistent with the result obtained by \citet{Chavanis_2013} using more cumbersome techniques (without perturbation theory and transition to the complex plane), but which allow one to obtain an eigenvalue not only for small but also for arbitrary collision frequencies.

\subsection{Eigenfunction Behavior in the Rare Collision Limit}
\label{sec:IIB}
We now analytically approximate the EF $g(u)$ near the resonance $u=c_r\equiv {\rm Re}(c)$ for small damping rates $\gamma\equiv -{\rm Im}(c) \ll 1$, representing it as $g(u)=h(u)\,\exp(-u^2)$. For $h(u)$ we obtain:
\begin{multline}
    \hspace{-8pt}(u-c)\,h+\frac{\alpha}{\sqrt{\pi}} u\hspace{-4pt}\int\limits_{-\infty}^{\infty}\hspace{-4pt} \d u\, h(u)\,\e^{-u^2} \hspace{-4pt}=
\frac{\mu}{2\i}\left(\frac{\d^2 h}{\d u^2}-2u\,\frac{\d h}{\d u}\right)\hspace{-2pt}.
\label{eq:h1}
\end{multline}

In the spirit of matched asymptotic expansions method \citep[e.g.,][]{Bar_Zel_71, Malley2014}, we divide the $u$-axis into a narrow inner region around the resonant level, $u=c_r$, and an outer region. We introduce the inner variable $U=(u-c_r)/{\ell}$ and set $\Gamma=\gamma/\ell$, where $\ell\equiv(\mu/2)^{1/3}\ll 1$ is the inner region scale. This same scale has previously appeared in weakly collisional plasma in the context of plasma echo \cite{Su_Oberman_1968} as well as in the context of Langmuir waves near resonance \cite{Auerbach_1977} (see also comments on this paper in the Conclusion).
Analogously, in shear flow stability theory of a nearly inviscid fluid, a viscous critical layer of width $\ell_\nu=\nu^{1/3}$ is introduced near the critical level $y_c$, where $U(y_c)=c$ \cite{Benney&Maslowe1975, Chu_Shu1987, ChSh1996, ChurShukh1996}.

With the normalization $\int_{-\infty}^{\infty} \d u\, h(u)\,\exp(-u^2)=1$, we obtain the approximate equation
\begin{equation}
\frac{\d^2 h}{\d U^2}-2 c_r\,\ell\,\frac{\d h}{\d U}-\i(U+\i \Gamma)\,h=\i\,\frac{\alpha}{\sqrt{\pi}}\,(U+c_r/\ell)
\end{equation}
with the asymptotic
$h \approx -{\alpha}/{\sqrt{\pi}}[1+(c_r/\ell-\i\Gamma)/{U}]$
as $|U|\to \infty$.
Letting $H(U)=h(U)+\alpha/\sqrt{\pi}$, we have $H={\cal O}(1/U)$ as $|U| \to \infty$, and
\begin{equation}
\frac{\d^2 H}{\d U^2}-2 c_r\,\ell\,\frac{\d H}{\d U}-\i (U+\i \Gamma)\,h=\i\frac{\alpha}{\sqrt{\pi}}\,\frac{c}{\ell}\,.
\label{eq:H}
\end{equation}
The Fourier transform,
\begin{equation}
{\tilde H}(q)=\frac{1}{2\pi}\int_{-\infty}^{\infty}\d U H(U)\,\e^{-\i qU}\,,
\end{equation}
yields the inhomogeneous first-order equation for the Fourier image ${\tilde H}(q)$:
\begin{equation}
\frac{\d{\tilde H}}{\d q}+(\Gamma-q^2 -2\,c_r\,\ell\,\i\,q)\,{\tilde H}=\i\,\frac{\alpha}{\sqrt{\pi}}\,\frac{c}{\ell}\,\delta(q),
\end{equation}
where $\delta(q)$ is the Dirac delta function. The solution is
\begin{equation}
{\tilde H}(q)=-\i\,\frac{\alpha}{\sqrt{\pi}}\,\frac{c}{\ell}\,\exp(q^3/3-\Gamma\,q +\i\,c_r\ell q^2)\, \Theta(-q)\,,
\end{equation}
where $\Theta(x)$ is the Heaviside step function. The inverse Fourier transform gives
\be
H(U)=-\i\,\frac{\alpha}{\sqrt{\pi}}\,\frac{c}{\ell} \int_0^{\infty}\d q\,\e^{-q^3/3-\i\,q\,(U+\i\Gamma -c_r\,\ell\, q)}\,.
\label{eq:H_plasma_1}
\ee
Finally, 
\begin{multline}
h(u)=-\frac{\alpha}{\sqrt{\pi}} - \i\,\frac{\alpha}{\sqrt{\pi}}\,\frac{c}{\ell} \\
\times
\int_0^{\infty} \d q\,\exp\{-\i\,q\,[(u-c)/\ell -c_r\,\ell\, q]\}\,\e^{-q^3/3}\,.
\label{eq:h_plasma_1}
\end{multline}
Integrating by parts verifies that this matches the outer solution $h(u)=-\alpha u/[\sqrt{\pi}(u-c)]$. Numerical calculations in the next section confirm this analytical approximation.

The analytical expression (\ref{eq:h_plasma_1}) for the inner-region EF $h(u)$, neglecting a small $\mathcal{O}(\ell)$ correction in the exponent, allows us to estimate the resonant region's half-width for $\Gamma \gtrsim 1$ (see Tab.\,\ref{tab1}). We find that the half-width, when expressed in the unscaled variable $u=U\ell$, depends mainly on the damping rate $\gamma$ (approximately $1.6\,\gamma\dots 2 \gamma$, as confirmed numerically) and weakly on $\ell$. We can also estimate the oscillation period and maximal relative amplitude.

Consider the exponential modulus in the integrand, $\exp(\Gamma q-\frac{1}{3}\,q^3)$, which peaks at $q=q_*\equiv\sqrt{\Gamma}$ with a value of $\exp(\frac{2}{3}\,\Gamma^{3/2})$. Expanding near the maximum, we have
\begin{equation}
\exp(\Gamma q-{\cas{1}{3}}\,q^3)=\exp \left[\,{\cas{2}{3}}\,\Gamma^{3/2}-\Gamma^{1/2}(q-q_*)^2\right].
\end{equation}
The integration over $q$ then yields (with $p=q-q_*$)
\begin{multline}
 I=\int_0^{\infty}\d q\,\e^{-\i\,q\,U}\,\e^{-q^3/3+\Gamma q}\approx \exp\left({\cas{2}{3}}\,\Gamma^{3/2}-\i q_*U\right) \\
 \times \int_{-q_*}^\infty \d p \,\exp\left(-\Gamma^{1/2} p^2-\i\,p\,U\right)\,.
\end{multline}
The lower limit, $p=-q_*=-\Gamma^{1/2}$, can be extended to $-\infty$ since the added interval's contribution is negligible for $\Gamma \gtrsim 1$. We have
\[
\int_{-\infty}^\infty \d p \,\exp (-\Gamma^{1/2} p^2- \i\,p\,U)
=\frac{\sqrt{\pi}}{\Gamma^{1/4}}\,\exp\Bigl(-\frac{U^2}{4\,\Gamma^{1/2}}\Bigr).
\]
Thus,
\be
h(U)\!=\!-\frac{\alpha}{\sqrt{\pi}}\Bigl[\frac{\i\,c}{\ell}\,\frac{\sqrt{\pi}}{\Gamma^{1/4}}\,\e^{ \case{2}{3}\,\Gamma^{3/2}-\case{1}{4}\,U^2\,\Gamma^{-1/2}-\i\,\sqrt{\Gamma}\,U}\!+\!1\Bigr].
\label{eq:h_res}
\ee
The expression (\ref{eq:h_res}) for $h(U)$ contains a factor $\exp(\frac{2}{3}\,\Gamma^{3/2}-\frac{1}{4}\,U^2 \Gamma^{-1/2})$ describing the envelope. This factor is large at $U=0$ and decreases exponentially with increasing $|U|$. We define the resonant region as the area around $U=0$ where this factor exceeds unity:
\begin{equation}
\frac{2}{3}\,\Gamma^{3/2}-\frac{U^2}{4\,\Gamma^{1/2}}>0.
\end{equation}
This yields a half-width of $\Delta U=\sqrt{{8}/{3}}\,\Gamma\approx 1.63\,\Gamma$, or
\begin{equation}
    \Delta u\approx 1.63\,\gamma
    \label{eq:hw}
\end{equation}
in ordinary variables $u=U\,\ell$ and $\gamma=\Gamma\,\ell$. This half-width is consistent with EF plots using a logarithmic scale for the ordinate axis. 
The coefficient 1.63 increases slightly when considering the pre-exponential factor $\alpha\, (c/\ell)\,\Gamma^{-1/4}$  in (\ref{eq:h_res}), which has a weaker dependence on $\mu$ (see Tab.\,\ref{tab1}, column  $\Delta u$ for $\alpha=9$, where one can observe how the half-width $\Delta u$ slowly approaches the value of $1.63\,\gamma$ from above as $\mu$ decreases). Note that the dependence of the limiting width of the resonance region on $\gamma$ alone (as well as its order of magnitude) was asserted in \citet{Ng1999}, although the authors could not confirm this numerically due to computational limitations that prevented calculations at sufficiently small $\mu$ values.


The oscillation period $T_u$ in $u$ is found from the exponent $\exp (-\i \sqrt{\Gamma} U)$:
\begin{equation}
T_u=2\pi\,\sqrt{\frac{\mu}{2\gamma}}.
\label{eq:per}
\end{equation}
From (\ref{eq:h_res}), the resonant peak height ${\mathfrak P}$ of the normalized EF $g(u)=h(u)\,e^{-u^2}$ is
\begin{equation}
\label{eq:peak}
{\mathfrak P}(\alpha,\mu)\approx
\frac{\alpha}{\Gamma^{1/4}}\,\frac{|c|}{\ell}\,\exp\Bigl({\case{2}{3}}\,\Gamma^{3/2}-c_r^2\Bigr),
\end{equation}
growing without limit as $\mu \to 0$.

\section{Numerical Validation}

For numerical evaluation of the eigenvalue problem at finite $\mu$, we expand in normalized Hermite polynomials $\brh_n(u)$, similar to \cite{Ng1999}:
\begin{equation}
g(u)=\e^{-u^2}\,h(u)=\e^{-u^2}\sum_{n=0}^\infty {A}_n \brh_n(u),
\label{eq:gu}
\end{equation}
where $\brh_n(u)={H_n}(u)/\bigl({2^{n/2} \sqrt{n!}\,\pi^{1/4}}\bigr)$ and
\begin{equation}
H_n(u)=(-1)^n\,\e^{u^2}\, \frac{\d^{\,n} }{\d u^n}(\e^{-u^2})
\end{equation}
are physicist's Hermite polynomials \citep[see, e.g.,][] {Gradshteyn_8}. For $n=0$, $\brh_0(u)=\pi^{-1/4}$, so normalizing $g(u)$ to unity yields ${A}_0=\pi^{-1/4}$ for all $\mu$. Equation (\ref{eq:h1}) leads to the linear system for expansion coefficients ${A}_n$:
\begin{align*}
    {A}_1 &=\sqrt{2}\,c\,{A}_0\,, \quad
       {A}_2 =(c+\i\mu)\,{A}_1-\frac{1+\alpha}{\sqrt{2}}\,{A}_0
       \,,\\
    {A}_{n+1} &=\sqrt{\frac{2}{n+1}}\left[(c+\i\mu\,n)\,{A}_n-\sqrt{\frac{n}{2}}\,{A}_{n-1}\right]
\end{align*}
for $n\ge 2$. Unlike \cite{Ng1999}, we formulate this as a standard matrix eigenvalue problem. Introducing ${\tilde c}=\sqrt{2}\,c$ and ${\tilde\mu}=\sqrt{2}\,\mu$, we obtain
\begin{equation}
{\tilde c}\,A_n=\sum\limits_{m=0}^\infty {\cal M}_{nm}({\tilde\mu})\, A_m, \ \  n\ge 0, 
\label{eq:matrix_eq}
\end{equation}
where ${\cal M}_{nm}(\tilde \mu)$ is a tridiagonal matrix:
\begin{align*}
{\cal M}_{nm}({\tilde\mu})\!=\!\left|
\begin{array}{cccccc}
0&\sqrt{1}&0&0&0&...\\
1\!+\!\alpha &-\i\,\tilde\mu&\sqrt{2}&0&0&...\\
0&\sqrt{2}&-2\,\i\,\tilde\mu&\sqrt{3}&0&...\\
0&0&\sqrt{3}&-3\,\i\,\tilde\mu&\sqrt{4}&...\\
0&0&0&\sqrt{4}&-4\,\i\,\tilde\mu&...\\
.&.&.&.&.&...\\
\end{array}\right| \,.
\end{align*}

To solve Eq. (\ref{eq:matrix_eq}), we used Matlab's sparse matrix function \texttt{eigs}, enhanced by Advanpix for multiprecision calculations. This finds a single eigenvalue near a preset value with quadrupole (128-bit) or higher precision. The matrix was truncated at some $n_{\rm max}$ for convergence.

\begin{table*}
\centering
\setlength{\tabcolsep}{6pt}
\begin{tabular}{cccccccccc}
\toprule
$\alpha$                 & $\lg\mu$ &$c_{\rm L}$        & $\Delta c/\mu$     & $\ell$&$\Gamma$& $\Delta u$ & ${\mathfrak P}$ & $T_u$ \\ 
\midrule
\multirow[t]{5}{*}{4.0}  & ---      &$2.00205-0.21688\i$&$-0.53996-0.56560\i$& ---   &---   &0.354 &---           & ---     \\[1pt]
                         & ${-4}$   &                   &\cellcolor{gray!20}$-0.53994-0.56561\i$&0.03685&5.8851&\cellcolor{gray!20}0.440 &34671         & 0.0954\\ 
\midrule
\multirow[t]{5}{*}{6.5}  & ---      &$2.29853-0.10924\i$&$-0.41198-0.68284\i$&---    &---   &0.178 & ---          &---    \\[1pt]
                         & ${-5}$   &                   &\cellcolor{gray!20}''&0.01711&6.3858&\cellcolor{gray!20}0.204 &131231         &0.0425\\
\midrule
\multirow[t]{5}{*}{9.0}  & ---      &$2.54582-0.05489\i$&$-0.29897-0.73744\i$& ---   &---   &0.0895& ---           &---        \\[1pt]
                         & ${-7}$   &                   &\cellcolor{gray!20}''&0.00369&14.891&\cellcolor{gray!20}0.0996&$2.10\;10^{17}$&0.005997\\  
                         & ${-6}$   &                   &\cellcolor{gray!20}''&0.00794&6.9124&\cellcolor{gray!20}0.1065&498278          &0.01896\\
                         & ${-5}$   &                   &\cellcolor{gray!20}$-0.29896-0.73744\i$&0.01711&3.2087&\cellcolor{gray!20}0.1193&71         &0.05997\\
                         
\bottomrule
\end{tabular}
\caption{Characteristics of the least-damped mode eigenfunction (EF) in the resonant region for a Maxwellian DF with varying $\alpha$ and $\mu$. Each block corresponds to an $\alpha$ value. The first row gives collisionless solution characteristics: Landau value $c_{\rm L}$, analytical collisional correction (\ref{eq:final_Omega_1}), parameters $\ell$, $\Gamma$, resonance half-width $\Delta u$ (\ref{eq:hw}), resonance peak ${\mathfrak P}$ (\ref{eq:peak}), and oscillation period $T_u$ (\ref{eq:per}). Subsequent rows show finite-$\mu$ quantities. Grey cells indicate numerical values; others use respective equations.}
\label{tab1}
\end{table*}

\begin{figure}[t!]
\centering
   \includegraphics[width=86mm, trim={5pt 10pt 5pt 8pt}, clip]{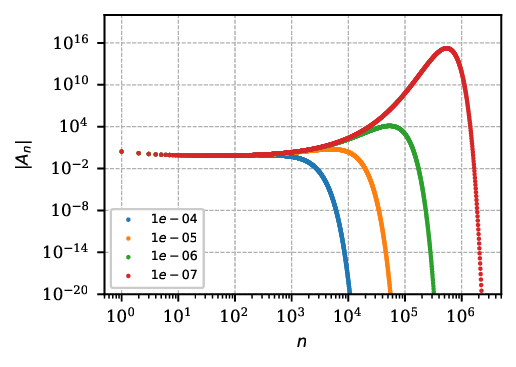}
   \caption{\footnotesize Modulus of expansion coefficients $|A_n|$, $n\ge 1$, for the least-damped mode's EF $g(u)$ with $\alpha=9$ at four $\mu$ values. $A_0=\pi^{-1/4}$ for all $\mu$. $|A_n|$ peaks at $n_* \approx \gamma/\mu$ for very low $\mu$.
   \vspace{-6pt}
}
\label{fig:a}
\end{figure}

\begin{figure*} [htbp]
\centering
   \includegraphics[width=145mm]{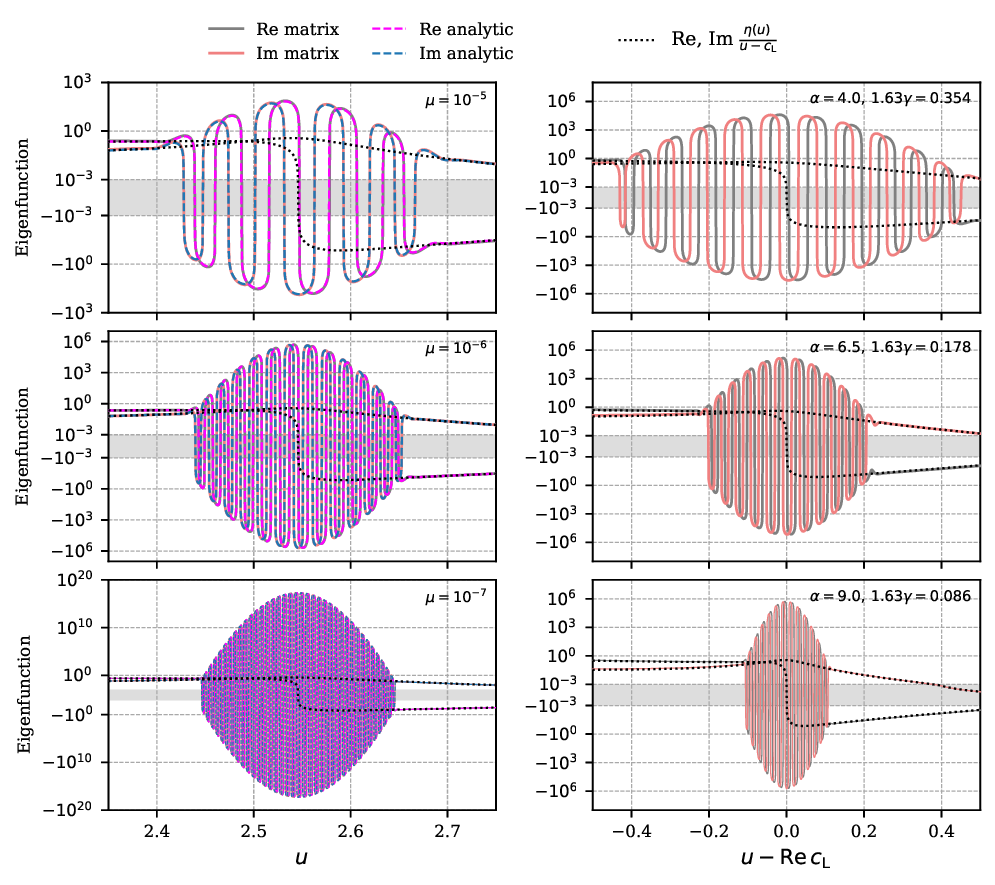}
   \vspace{-3mm}
   \caption{\footnotesize Least-damped mode eigenfunctions near resonance $u={\rm Re}\;c$. Left: EF deformation for $\alpha=9$ as $\mu$ decreases (solid: numerical solutions from (\ref{eq:matrix_eq}), dashed: analytical approximation (\ref{eq:H_plasma_1}), dotted: nonresonant solution (\ref{eq:g_0})). Right: Resonance range width dependence on $\gamma$ for three $\alpha$ values, supporting our analytical result (\ref{eq:hw}). EFs are plotted using symlog scaling (logarithmic with linear scaling near zero, indicated by grey shading).
}
\label{fig:ef}
\end{figure*}

Numerical validation is presented in Tab.\,\ref{tab1} and Figs.\,\ref{fig:a}, \ref{fig:ef}. Tab.\,\ref{tab1} shows damping solution characteristics for Maxwellian DF with varying $\alpha$ and $\mu$. The first row of each block gives collisionless Landau eigenfrequencies $c_{\rm L}$ and collisionless corrections from (\ref{eq:DEq1}) and (\ref{eq:final_Omega_1}). Subsequent rows show finite-$\mu$ quantities. Grey cells mark numerical estimates of analytical counterparts: $\Delta c/\mu$ as the difference between $c$ for given $\mu$ and $c_{\rm L}$, and $\Delta u$ as the half-width between the the EF's imaginary part rightmost and leftmost zeros. One can observe that $\Delta c/\mu$ is nearly constant for considered $\mu$, indicating that $c$ smoothly transits to the collisionless Landau eigenfrequencies $c_{\rm L}$ as $\mu\to 0$. Also, the half-width approaches its analytical value (\ref{eq:hw}) in this limit.

Unlike eigenfrequencies, the EFs do not converge to any EF in the collisionless limit. This is already evident in the lack of convergence at $\mu=0$, indicated by the growth of $|A_n|$ from solving Eq. (\ref{eq:matrix_eq}) (Fig.\ref{fig:a}). Numerically, maxima are attained for $n=n_* < \gamma/\mu$, with $n_*$ approaching $\gamma/\mu$ as $\mu\to 0$. Eigenvalue problem convergence requires $n_{\rm max}$ several times larger than $n_*$.
As $\mu\to 0$, the maximum shifts to larger $n_*$, reflecting the divergence of the series for EF and, as a consequence, its absence in the collisionless case.

Figure \ref{fig:ef} validates our analytical EF expression (\ref{eq:H_plasma_1}), which closely matches the numerical matrix solution (\ref{eq:gu}, \ref{eq:matrix_eq}). The left column (for $\alpha=9$ and decreasing $\mu$) shows increasingly oscillatory functions in $u$, with smaller oscillation periods and increasing resonance peaks. The resonance half-width decreases slightly, approaching the analytical limit (\ref{eq:hw}). Outside the resonance region, the solution matches the nonresonant solution (\ref{eq:g_0}). The right column shows EFs for three $\alpha$ values with damping rates in a 4:2:1 ratio and $\mu$ values of $10^{-4}$, $10^{-5}$, and $10^{-6}$, respectively. The resonance region width generally follows our analytical finding (\ref{eq:hw}), although the estimate's accuracy is limited for the upper model due to the relatively large $\ell$ (Tab.\,\ref{tab1}).

\section{Conclusion}

Nearly three-quarters of a century after Landau's discovery, questions about its interpretation and applications remain highly relevant across physics. As detailed in Ryutov's review \cite{Ryutov_1999} commemorating the 50th anniversary of Landau's famous publication, Landau damping extends far beyond plasma physics into stellar dynamics \citep{FP_1984}, hydrodynamics \citep{Miles1957}, and even quark-gluon plasma \citep{Baier1992}. Despite the significant differences between these media, there exists a remarkable analogy in how Landau damping manifests in their inherent oscillations. A fundamental question arises in all cases: how do nearly vanishing dissipative factors affect Landau damping? For small amplitude oscillations, this leads to questions about the behavior of eigenvalues and eigenfunctions (EFs) as dissipation approaches zero.

The series of papers \cite{Ng1999, Ng2004, Ng2021} as well as \cite{Auerbach_1977} analyze the transition of collisional modes in idealized scenarios of infinite, homogeneous, isotropic plasma with a Maxwellian DF and its stellar dynamical counterpart in the collisionless limit. These studies show that eigenvalues transition smoothly to discrete Landau frequencies as collisions vanish, rather than to the real continuous spectrum of singular van Kampen modes. However, despite presenting EFs for relatively small collision frequencies $\mu$, these works examine values that are still too large to reveal what truly happens to the EFs as the collisionless limit is approached. This gap was addressed in our recent work \citep{2025AJ....169..224P} on Landau damping in self-gravitating stellar systems, where we pushed numerical methods to examine collision frequencies as low as $\mu=10^{-10}$. This allowed us to study in unprecedented detail how the EF evolves and ultimately disappears in the strictly collisionless limit.

 In this paper, using the least-damped mode, we demonstrate that despite the smooth transition of eigenvalues to the Landau collisionless value $c_{\rm L}$, a true EF becomes increasingly oscillatory as $\mu$ approaches zero, with both the oscillation frequency and amplitude diverging in the strict $\mu=0$ limit. Building on our analysis of collisionless infinite homogeneous self-gravitating media \cite{PSB}, we show that the DF perturbation with self-similar decay exists only on complex-valued $u$-contours passing below $c_{\rm L}$, not for real-valued velocities. This insight enables us to derive refined Landau eigenvalues, $c=c_{\rm L}+\Delta c(\mu)$, and approximate analytical expressions for the EF at very small but finite $\mu$. We analyze its behavior near resonance, deriving expressions for the resonance region's half-width, peak value, and oscillation period in $u$. Notably, we find the resonance region maintains a finite limiting half-width of $1.63\gamma$ even as $\mu$ approaches zero.

We demonstrate numerically that the problem of determining the eigenvalues and EFs can be reduced to a standard linear eigenvalue problem for a tridiagonal matrix. Utilizing efficient computational techniques, we can solve this problem for matrices of rank on the order of $10^9$ on a standard desktop computer, enabling the investigation of models with collision frequencies as low as $\mu \sim 10^{-9}$. These high-resolution calculations provide an excellent opportunity to validate our theoretical findings. The comparison of the calculations and analytical solutions reveals strong agreement.

During the review process, our attention was drawn to significant earlier analytical work by \citet{Auerbach_1977} that examined DF behavior near resonance in weakly collisional plasma. Auerbach also introduced the $\mu^{1/3}$ scaling for the inner region using the same matched asymptotic expansions method \citep{Bar_Zel_71},  widely applied in high Reynolds number shear flow hydrodynamics (e.g., \citep{Haberman_1972, Huerre1980,Chu_Shu1987}).

Our analysis and high-precision numerical calculations reveal important aspects that Auerbach's analysis did not capture. We specifically clarify the behavior of the resonance region width. When $\mu$ is not very small, $\mu^{1/3}$ is comparable to the damping rate $\gamma$, making it difficult to determine the resonance region's true extent. For proper matching of inner and outer solutions, an intermediate scale $L$ must exist where $\ell \ll L \ll 1$ (with $\ell \equiv (\mu/2)^{1/3}$ being the inner region scale). When parameter $\Gamma \equiv \gamma/\ell$ becomes of order unity or larger, the resonant region extends beyond $\ell$ into the intermediate region, maintaining a finite limiting width of $1.63\,\gamma$ as $\mu$ approaches zero, rather than vanishing. This finding is derived in our analytic expressions (Eqs. \ref{eq:h_res}--\ref{eq:hw}) and confirmed numerically. Moreover, our analytical expressions for the EF (Eq. \ref{eq:h_plasma_1}) and peak estimate (Eq. \ref{eq:peak}) fit the exact numerical solutions more accurately than Auerbach's counterparts (Eqs. 24 and 25), particularly in correctly reproducing the complex phase of the resonant solution. Through this combined analytical and numerical approach, we provide a complete description of the resonance region, including its width, peak amplitude, oscillation frequency, and the EF with correct phase structure.

We do not pursue here the delicate question of the Lenard-Bernstein collision term's applicability for studying weak collision effects on Landau damping of small-amplitude Langmuir waves. For discussions of this issue, see \cite{LenBern, Su_Oberman_1968, Auerbach_1977, Ng1999, Chavanis_2013}.

A more sophisticated collision term was employed by \citet{Callen_2014}, accounting for the three-dimensional nature of Coulomb collisions. This approach produces not only diffusion in $|\bv|$ but also angular scattering, resulting in an effective collision frequency $\nu_{\rm eff}$ that, as shown in \cite{Callen_2014}, exceeds the ordinary collision frequency $\nu$. This more complex collision term does not change the main result in the long-time asymptotic regime ($t \gg 1/\nu_{\rm eff}$) -- the electric field still dampens at Landau's rate. However, when solving the initial disturbance evolution problem (following Landau's approach), it allows tracking the DF's behavior during intermediate stages ($t\ll 1/\nu_{\rm eff}$ and $t \gtrsim 1/\nu_{\rm eff}$). Unfortunately, such a sophisticated collision term is too complex for the eigenvalue problem we consider here.

\bigskip

\begin{acknowledgments}
We thank M. Weinberg for pointing out the papers by Ng, Bhattacharjee \& Skiff,  and are grateful to an anonymous referee who brought to our attention the paper by Auerbach, which has some overlap with our Sec.\,\ref{sec:IIB}. This research was partially supported by NSF grant PHY-2309135 to the Kavli Institute for Theoretical Physics (E. Polyachenko), and by the Russian Academy of Sciences Program No. 28 (subprogram II, 'Astrophysical Objects as Cosmic Laboratories') and the Ministry of Science and Higher Education of the Russian Federation (I. Shukhman). High-precision calculations were performed with the Advanpix Multiprecision Computing Toolbox.
\end{acknowledgments}


%
%

%



\bibliography{main}{}
\bibliographystyle{apsrev4-1}

\end{document}